\documentclass
[aps,prl,superscriptaddress,showpacs,twocolumn,footinbib]{revtex4-1}
\usepackage{amsmath,amsfonts,amssymb}
\usepackage{graphicx,float,calc}
\usepackage{color,bm}
\usepackage[colorlinks,urlcolor=blue,citecolor=blue,linkcolor=blue]{hyperref}
\usepackage{amsmath}
\usepackage{amsfonts}
\usepackage{amssymb}
\usepackage{graphicx}
\usepackage[sort&compress]{natbib}%
\providecommand{\U}[1]{\protect\rule{.1in}{.1in}}

\begin{document}
\author{Youjiang Xu}
\author{Han Pu}
\affiliation{Department of Physics and Astronomy, and Rice Center for Quantum Materials,
Rice University, Houston, Texas 77251-1892, USA}
\title{Emergent Universality in a Quantum Tricritical Dicke Model}

\begin{abstract}
We propose a generalized Dicke model which supports a quantum tricritical
point. We map out the phase diagram and investigate the critical behaviors of
the model through exact low-energy effective Hamiltonian in the thermodynamic
limit. As predicted by the Landau theory of phase transition, the order
parameter shows non-universality at the tricritical point. Nevertheless, as
a result of the separation of the classical and the quantum degrees of
freedom, we find a universal relation between the
excitation gap and the entanglement entropy for the entire critical line
including the tricritical point. Here the universality is carried by the
emergent quantum modes, whereas the order parameter is determined classically.

\end{abstract}
\maketitle

\emph{Introduction ---} Tricritical point was first proposed by Griffiths
within the Landau theory of phase transition \cite{PhysRevLett.24.715}. A
tricritical point is where ordinary critical manifolds intersect \cite{PhysRevB.8.346}. In the physically accessible phase diagram, it can appear as a point where a
first-order phase transition boundary and a second-order one meet
\cite{PhysRevLett.24.715,PhysRevB.8.346}. As for the critical behaviors, the tricritical point
normally belongs to a universality class different from that of other points
on the critical line \cite{PhysRevLett.28.675,henkel2013conformal}.

Quantum phase transition\cite{sachdev2011quantum} has been under intensive study over many years, and
is a central subject in the study of numerous important solid state materials
such as high temperature superconductors and heavy fermions. Systems that support quantum tricritical point
(QTP) are, however, very rare. Recently it has been found that QTP exists in
certain magnetic materials \cite{PhysRevB.97.045139,friedemann2018quantum}. In the present work, we
construct a generalized Dicke model which not only supports a QTP, but that
the QTP exhibits a special feature: Despite the non-universal critical
exponent that distinguishes the QTP from other critical points, there exists a
universal relation between the excitation gap and the entanglement entropy of
the system, which applies to all the critical points of the model. This
universal relation characterizes the quantum fluctuations and the emergent
collective modes of the model.

The Dicke model \cite{PhysRev.93.99,Garraway1137} describes an ensemble of
two-level systems interacting with a quantized bosonic mode. Though
originated as a model of atom-light interaction, the Dicke model can be
realized in various experimental settings, including quantum gases
\cite{PhysRevA.75.013804,PhysRevLett.104.130401,baumann2010dicke,PhysRevA.97.043858}%
, superconducting circuit
\cite{lamata2017digital,mezzacapo2014digital,langford2017experimentally}, and
solid state systems \cite{Li794}. The Dicke model features the famous
superradiant phase transition \cite{PhysRevA.7.831}, where the
bosonic mode becomes macroscopically occupied if the atom-light interaction
strength exceeds a threshold value and the system enters the superradiant
phase. While the ground-state phase diagram can be determined classically through a
mean-field approach, the superradiant phase transition is associated with a divergent entanglement
entropy \cite{PhysRevLett.92.073602,PhysRevA.71.053804} which suggests non-trivial
effects induced by quantum fluctuations. In the generalized Dicke Hamiltonian
we study in this work, defined in Hamiltonian (\ref{H}) below, an additional
dimension is present, such that the generalized model
extends the critical point in the Dicke model into a line and the
second-order superradiant phase transition can be tuned into a first-order one
across a QTP. As a consequence, we shall call the model under study the
\emph{quantum tricritical Dicke model}. We will explore the phase diagram and
the critical behavior of this model at zero temperature in the thermodynamic limit.

\begin{figure}
[ptb]\centering
\includegraphics[width=0.45\textwidth]{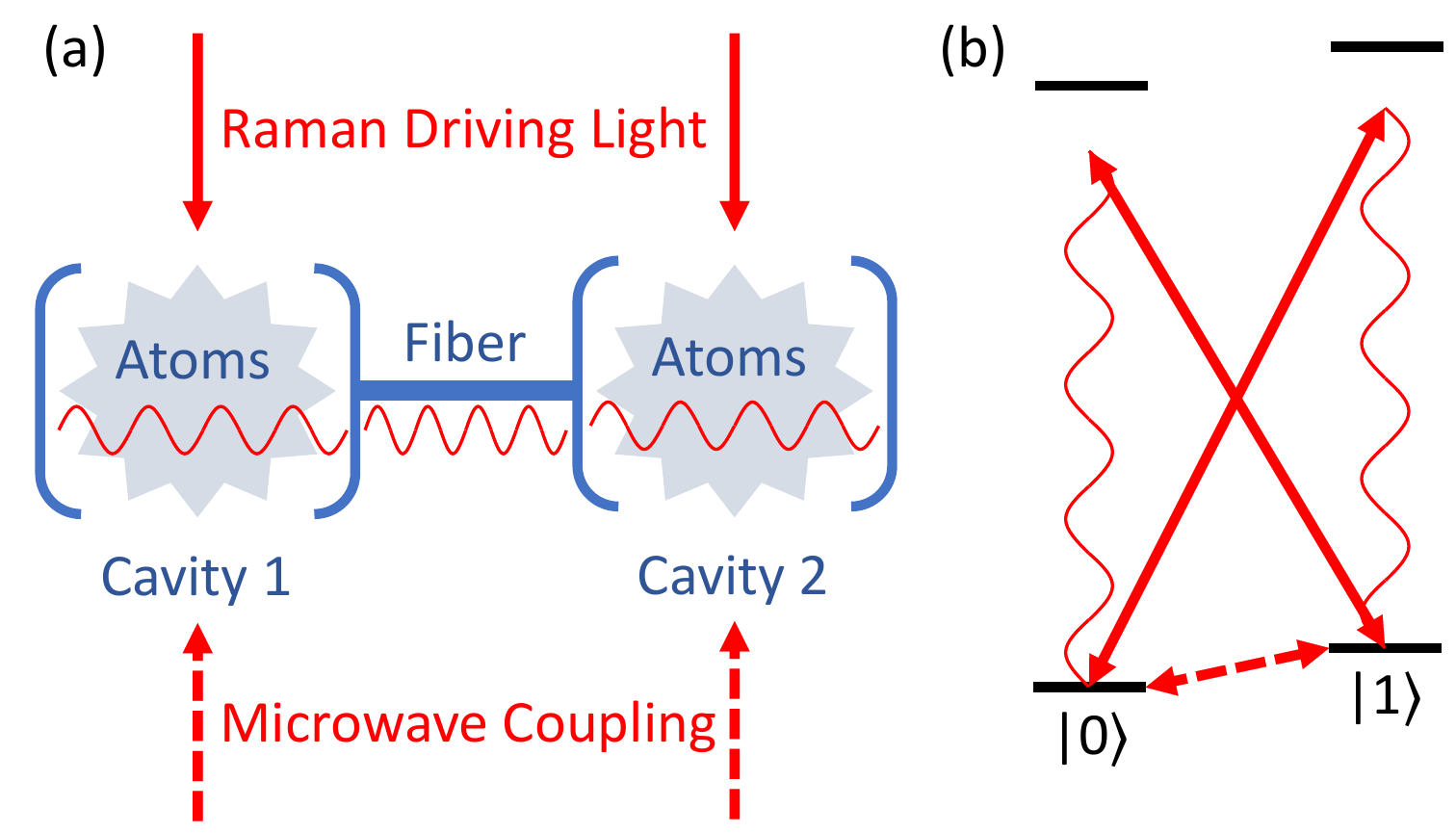}\caption{(a) A schematic representation of potential realization of the tricritical Dicke model. We have two identical cavities linked by an optical fiber and atoms are trapped within each cavity. We assume that the fiber coupled cavity system contains one normal mode that is near resonant with the atomic transition and all other modes can be neglected.  (b) An external light, together with the dominant cavity mode, drives a Raman transition between two low-energy states labelled as $\left\vert 0\right\rangle $ and $\left\vert 1\right\rangle $, which realizes the Dicke coupling as proposed in \cite{PhysRevA.75.013804}. In addition, a microwave field directly couples the two spin states. The two microwaves for each cavity have a phase difference of $\pi$, and serve as the effective Zeeman field in Eq.~(\ref{HSB}).   }\label{experiment}
\end{figure}

\emph{Model ---} The quantum tricritical Dicke model is obtained by partially
breaking the exchange symmetry between the two-level atoms in the Dicke Hamiltonian
$H_{\mathrm{Dicke}}$ through an additional term $H_{\mathrm{SB}}$
\begin{align}
H  &  =H_{\mathrm{Dicke}}+H_{\mathrm{SB}}\,,\label{H}\\
H_{\mathrm{Dicke}}  &  =\omega b^{\dag}b+\sum_{i=1}^{N}\left[  \frac{\delta
}{2}\sigma_{i}^{\left(  z\right)  }+\frac{g\left(  b+b^{\dag}\right)  }%
{2\sqrt{N}}\sigma_{i}^{\left(  x\right)  }\right]  \,,\\
H_{\mathrm{SB}}  &  =\frac{\varepsilon}{2}\sum_{i=1}^{N}\left(  -1\right)
^{i}\sigma_{i}^{\left(  x\right)  }\,.\label{HSB}
\end{align}
Here the operator $b$ represents the annihilation operator for the bosonic
light mode, $\sigma_{i}$'s are Pauli matrices describing the $i^{\rm th}$ atom.
$\omega,\delta$ and $g$ represent the light frequency, the atom excitation
energy, and the atom-light interaction strength, respectively. Without loss of
generality, all these parameters are taken to be non-negative. In
$H_{\mathrm{Dicke}}$, all atoms are identical. This symmetry is, however,
broken by $H_{\mathrm{SB}}$ which separates the atoms into two groups: one
group experiences an effective Zeeman field along the $x$-axis, while the
other group sees the Zeeman field in the opposite direction. We choose the
total number of atoms $N$ to be even. As we will see, the
second-order quantum phase transition in the conventional Dicke model can be
tuned into a first-order one by increasing the strength $\varepsilon$ of the
symmetry breaking term. In Fig.~\ref{experiment}, we present a potential experimental realization of our model, which involves Raman transition \cite{PhysRevA.75.013804} in two cavities linked by optical fiber \cite{Pyrkov_2013,PhysRevA.98.043616}. If $N=1$, our model reduces to the asymmetric Rabi model \cite{1751-8121-50-17-174001}, which has received much attention recently, partially due to its relevance in circuit QED \cite{niemczyk2010circuit}.

To proceed, we carry out a series expansion of the Hamiltonian in terms of
$1/N$, so that a solvable low-energy effective Hamiltonian can be obtained. To
this end, we introduce the shifted bosonic operator $b_{1}\equiv b-\psi$. Here
$\psi$ is a $c$-number, which can be regarded as arbitrary for now. After
rotating the Pauli matrices, we can recast the Hamiltonian into the following
form
\begin{align*}
H  &  =\omega_{1}b_{1}^{\dag}b_{1}+\omega_{1}\psi\left(  b_{1}+b_{1}^{\dag
}\right)  +\omega_{1}\psi^{2}\\
&  +\sum_{i,\mathrm{even}}\left[  \frac{\omega_{2}}{2}\sigma_{i}^{\left(
z\right)  }+\frac{g\left(  b_{1}+b_{1}^{\dag}\right)  }{2\sqrt{N}}\left(
\sin\theta_{2}\sigma_{i}^{\left(  z\right)  }+\cos\theta_{2}\sigma
_{i}^{\left(  x\right)  }\right)  \right] \\
&  +\sum_{i,\mathrm{odd}}\left[  \frac{\omega_{3}}{2}\sigma_{i}^{\left(
z\right)  }+\frac{g\left(  b_{1}+b_{1}^{\dag}\right)  }{2\sqrt{N}}\left(
\sin\theta_{3}\sigma_{i}^{\left(  z\right)  }+\cos\theta_{3}\sigma
_{i}^{\left(  x\right)  }\right)  \right]  ,
\end{align*}
where
\begin{align*}
\omega_{1}  &  \equiv\omega\,,\\
\omega_{2,3}  &  \equiv\sqrt{\delta^{2}+\left(  {2g\psi}/\sqrt{N}%
\pm\varepsilon\right)  ^{2}}\,,\\
\theta_{2,3}  &  \equiv\tan^{-1}[({{2g\psi}/\sqrt{N}\pm\varepsilon})/{\delta
}]\,.
\end{align*}
We then define two collective atomic angular momentum operators for the two
groups of atoms:
\[
J_{2}^{\left(  x,y,z\right)  }\equiv\frac{1}{2}\sum_{i,\mathrm{even}}%
\sigma_{i}^{\left(  x,y,z\right)  }\,,\ \;\;J_{3}^{\left(  x,y,z\right)
}\equiv\frac{1}{2}\sum_{i,\mathrm{odd}}\sigma_{i}^{\left(  x,y,z\right)  }\,.
\]
Without loss of generality, we restrict the Hilbert space to the subspace with maximum
$J_{2}$ and $J_{3}$. These operators can be represented by two new bosonic operators $b_{2},b_{3}$ by means of the Holstein-Primakoff mapping
\cite{PhysRev.58.1098}:
\[
J_{i}^{\left(  z\right)  }=b_{i}^{\dag}b_{i}-N/4\,,\;\;\;\;J_{i}^{\left(
+\right)  }=b_{i}^{\dag}\sqrt{N/2-b_{i}^{\dag}b_{i}}\,,\;\;\;i=2,3\,.
\]
By expanding $J_{i}^{\left(  \pm\right)  }$ in powers of $1/N$, the following
effective Hamiltonian of $H$ can be constructed:
\begin{align}
H_{\text{\textrm{eff}}}  &  =\omega_{1}\left(  b_{1}^{\dag}b_{1}+\psi
^{2}\right)  -N(\omega_{2}+\omega_{3})/4\nonumber\\
&  +\left[  \omega_{1}\psi-{g\sqrt{N}\left(  \sin\theta_{2}+\sin\theta
_{3}\right)  }/{4}\right]  \left(  b_{1}+b_{1}^{\dag}\right) \nonumber\\
&  +\sum_{i=2,3}\left[  \omega_{i}b_{i}^{\dag}b_{i}+\frac{g\cos\theta_{i}%
}{2\sqrt{2}}\left(  b_{i}+b_{i}^{\dag}\right)  \left(  b_{1}+b_{1}^{\dag
}\right)  \right]  . \label{Heff}%
\end{align}
We label the set of states satisfying $\langle b_{i}^{\dag}b_{i}%
\rangle=o(N),i=2,3$ as $V$, and $H-H_{\text{\textrm{eff}}}%
=o(H_{\text{\textrm{eff}}})$ holds only in $V$ when $N\rightarrow\infty$.
$H_{\text{eff}}$ is quadratic and solvable for arbitrary $\psi$. However, if
we want $V$ to contain the low-energy states of $H$ and $H_{\text{\textrm{eff}%
}}$, the second line in Eq. (\ref{Heff}) is necessarily small. This can be achieved by choosing $\psi$ to coincide with the expectation value $\left\langle
b\right\rangle $, which can be identified as the order parameter in the mean-field theory, as we show below.

The mean-field order parameter minimizes the
dimensionless mean-field energy-per-atom functional \footnote{
	Following the standard
	procedure, we obtain the mean-field Hamiltonian $H_{\mathrm{MF}}$ by replacing
	the bosonic operator $b$ in the Hamiltonian $H$, Eq.~(\ref{H}) with its expectation value \unexpanded{$ \varphi\equiv\langle b \rangle$}
: $H_{\mathrm{MF}}(\varphi
	)=\omega\varphi^{2}+\sum_{i=1}^{N}\mathbf{B}_{i}\cdot{\boldsymbol{\sigma}_{i}%
	}$, with $\mathbf{B}_{i}=(\Delta/2)\hat{z}+[g\varphi/\sqrt{N}+(-1)^{i}%
	\varepsilon/2]\hat{x}$. The mean-field ground state is reached when the $i$-th atom is
	polarized along the direction anti-parallel to $\mathbf{B}_{i}$, with the corresponding
	mean-field energy functional $h_{\mathrm{MF}}(\varphi)=\omega\varphi
	^{2}-\sum_{i=1}^{N}|\mathbf{B}_{i}|$, which yields $f(z)$ in Eq.~(\ref{fz}) of the text.
}:%
\begin{equation}
f\left(  z\right)  =\frac{z^{2}/y-\sqrt{1+2xz+z^{2}}-\sqrt{1-2xz+z^{2}}}{2}\,,
\label{fz}%
\end{equation}
where $x\equiv\varepsilon/\omega_{0}$ and $y\equiv g^{2}/(\omega\omega_{0})$
are two dimensionless system parameters with $\omega_{0}\equiv\sqrt
{\varepsilon^{2}+\delta^{2}}$, and
\begin{equation}
z=2g\left\langle b\right\rangle /(\omega_{0}\sqrt{N}\,),
\end{equation}
is the normalized order parameter. As a result, the coefficient of the term
linear in $b_{1}$ and $b_{1}^{\dagger}$ in Eq.~(\ref{Heff}) vanishes since%
\begin{equation}
{\omega_{1}}\psi-{g\sqrt{N}\left(  \sin\theta_{2}+\sin\theta_{3}\right)  }%
/{4}={\sqrt{N}gf^{\prime}(z)}/{2}=0\,. \label{cond}%
\end{equation}
Consequently, the eigenstates of $H_{\text{\textrm{eff}}}$ satisfies $\langle
b_{1}\rangle=0$, which self-consistently yields $\psi=\left\langle
b\right\rangle $.

\emph{Low-energy effective Hamiltonian and phase diagram ---} With $\psi$
given by the mean-field theory, $H_{\text{\textrm{eff}}}$ becomes%
\begin{align}
H_{\text{\textrm{eff}}} &  =H_{\mathrm{C}}+H_{\mathrm{Q}}\,,\label{CQ}\\
H_{\mathrm{C}} &  =\frac{N\omega_{0}}{2}f\left(  z\right)  ,\label{Hc}\\
H_{\mathrm{Q}} &  =\sum_{i=1,2,3}\omega_{i}b_{i}^{\dag}b_{i}+\sum_{i=2,3}%
\frac{g\cos\theta_{i}}{2\sqrt{2}}\left(  b_{1}+b_{1}^{\dag}\right)  \left(
b_{i}+b_{i}^{\dag}\right)  .\label{HQ}%
\end{align}
If we regard $z$ as a classical degree of freedom when we search for the
ground state of $H_{\text{\textrm{eff}}}$ in Eq. (\ref{CQ}), then by taking
the thermodynamic limit, the classical degree of freedom becomes fully
separated from the quantum ones, in the sense that $H_{\mathrm{Q}}=o\left(
H_{\mathrm{C}}\right)  $ when $N\rightarrow\infty$. As a result, $z$ is fully
determined by the classical part $H_{\mathrm{C}}$, independent from the
quantum part $H_{\mathrm{Q}}.$ The separation of the two kinds of degrees of
freedom contributes to the emergence of a new universality as we will show when we discuss the critical behavior of the model.

\begin{figure}
[ptb]\centering
\includegraphics[width=0.45\textwidth]{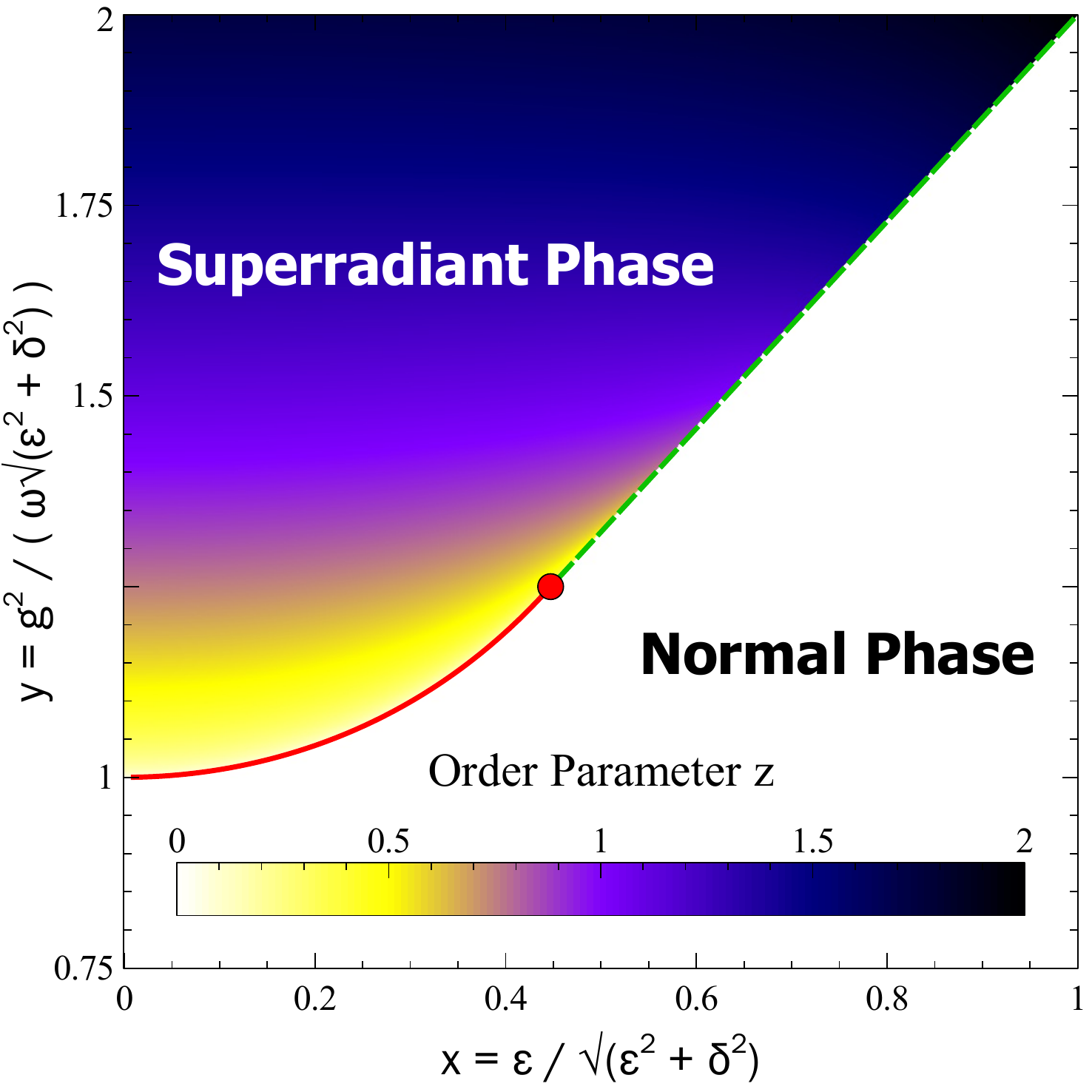}\caption{The phase diagram of the
tricritical Dicke model. The order parameter $z$ vanishes in
the normal phase and is finite in the superradiant phase. The quantum tricritical
point (QTP) is marked by a red dot, which is located at the intersection of the
second-order phase transition boundary (red solid line) and the first-order phase
transition boundary (green dashed line).  }\label{phasediagram}
\end{figure}

By minimizing $H_{\mathrm{C}}$, we obtain the order parameter $z$, from which
we can map out the phase diagram \footnote{Writing
	$f\left(  z\right)  $ as a power series of $z$: $f=\sum_{n=0}^{\infty}\frac{c_{2n}}{\left(  2n\right)  !}z^{2n}$,
we can extract the coefficients of the series and analytically obtain the equations which determines phase boundary and the tricritical point.
For example, the critical line (2nd-order phase transition boundary) is determined by $c_2=0$ which yields $y_{c}=\left(  1-x_{c}^{2}\right)  ^{-1}$.
The QTP is determined by $c_2=c_4=0$, which yields Eq.~(\ref{tricritical}) of the text.} in the $xy$-parameter space as shown in
Fig.~\ref{phasediagram}. The normal and the superradiant phases are
characterized by $z=0$ and $z>0$, respectively. The entire phase boundary is split into a solid line and
a dashed line, which mark the 2nd- and the 1st-order phase transition,
respectively. These two lines join together at the QTP marked as a red dot in
the figure. The position of the QTP is given by
\begin{equation}
\left(  x_{tc},\;y_{tc}\right)  =\left(  1/\sqrt{5},\;5/4\right)  \,.
\label{tricritical}%
\end{equation}
The presence of the QTP is one of the main results of our work.

While $H_{\mathrm{C}}$ determines the order parameter, $H_{\mathrm{Q}}$ in
Eq.~(\ref{HQ}) gives the quantum fluctuation above the ground state, from
which we can find the excitation gap and the ground state atom-light
entanglement entropy. It is convenient to define the generalized position and
momentum operators as
\[
X_{i}=\frac{b_{i}+b_{i}^{\dag}}{\sqrt{2\omega_{i}}},\,\;\;\;\;P_{i}%
=\sqrt{\frac{\omega_{i}}{2}}\frac{b_{i}-b_{i}^{\dag}}{i}\,,\;\;\;i=1,2,3\,,
\]
in terms of which, $H_{\mathrm{Q}}$ takes the form of a Hamiltonian
that describes a 3-dimension harmonic oscillator:
\begin{align}
H_{\mathrm{Q}}  &  =\frac{1}{2}{\sum_{ij}P_{i}^{2}+\frac{1}{2}\left(
\mathbf{\Omega}^{2}\right)  _{ij}X_{i}X_{j}-}\frac{{\omega_{i}}}{{2}%
}\,,\label{HQ3}\\
\mathbf{\Omega}^{2}  &  \mathbf{\equiv}%
\begin{pmatrix}
\omega_{1}^{2} & \lambda_{12} & \lambda_{13}\\
\lambda_{12} & \omega_{2}^{2} & 0\\
\lambda_{13} & 0 & \omega_{3}^{2}%
\end{pmatrix}
\,,\;\;\lambda_{ij}\equiv\sqrt{\frac{\omega_{i}\omega_{j}}{2}}g\cos\theta
_{j}\,.\nonumber
\end{align}
Here $X_{1}$ and
$P_{1}$ represent the original photonic degrees of freedom, while $X_{2,3}$ and $P_{2,3}$ represent the atomic degrees of freedom.

\begin{figure}
[ptb]%
\centering\includegraphics[width=0.45\textwidth]{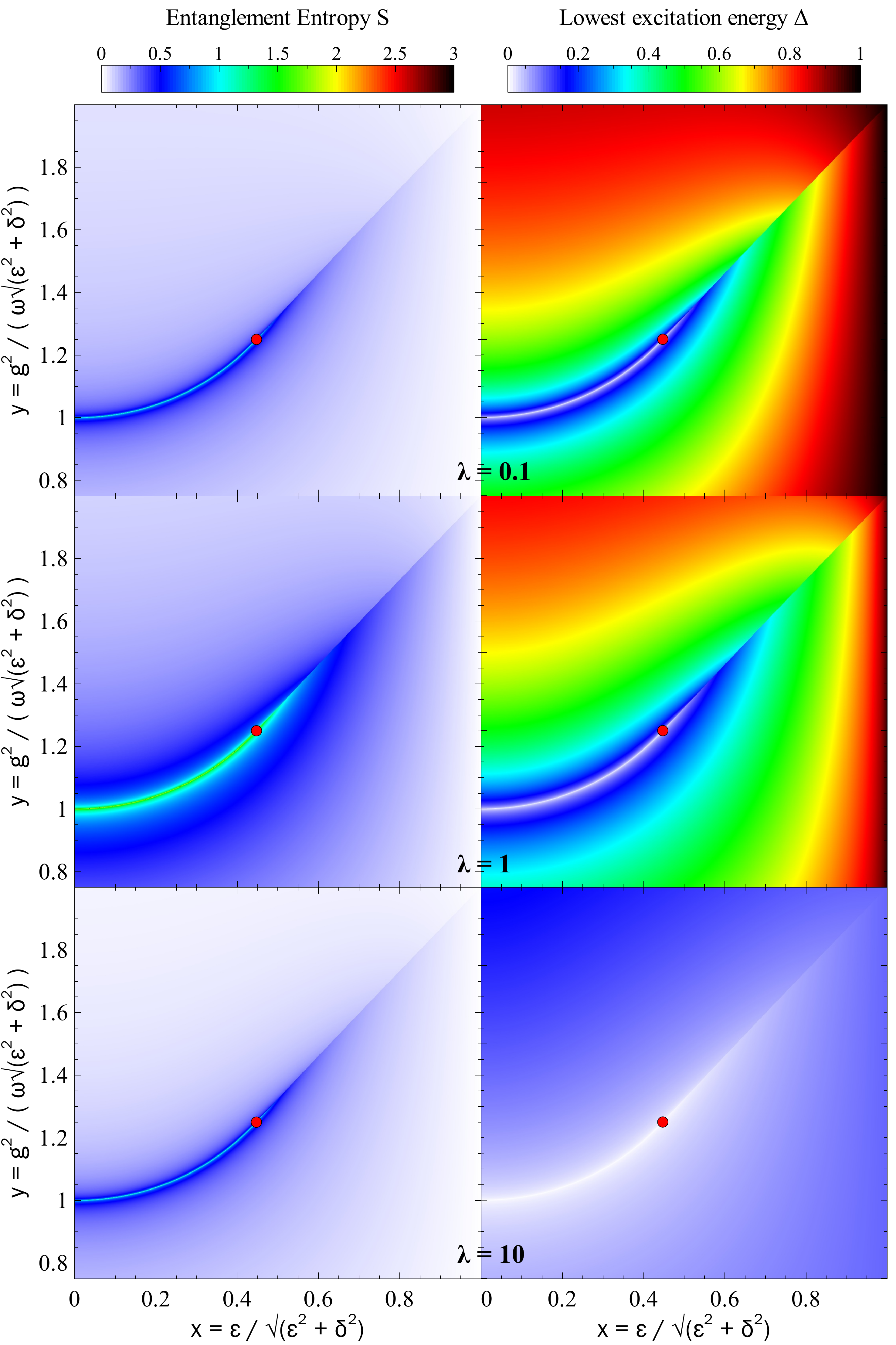}\caption{The atom-light entanglement entropy $S$ (left panel) and the lowest excitation
energy $\Delta$ (right panel) as functions of $x$ and $y$ for $\lambda \equiv \omega/\omega_0=0.1$, 1, 10 (from top to bottom). The QTP
is marked by the red dot as in Fig.~\ref{phasediagram}.}\label{entropygap}
\end{figure}

From Hamiltonian (\ref{HQ3}), it follows that the lowest excitation energy,
i.e., the excitation gap, $\Delta$, is given by the smallest eigenvalue of
$\mathbf{\Omega}$, and the ground state wave function $\Psi_{G}$ is a Gaussian
of the form%
\begin{equation}
\Psi_{G}\left(  \mathbf{X}\right)  =\left(  \frac{\det\mathbf{\Omega}}{\pi
^{3}}\right)  ^{1/4}\exp\left(  -\frac{\Omega_{ij}X_{i}X_{j}}{2}\right)  \,,
\end{equation}
from which we can calculate the reduced density matrix of the light field by
integrating out the atomic degrees of freedom:
\begin{equation}
\rho\left(  X_{1},X_{1}^{\prime}\right)  =C\exp\left(  -\frac{1}{2}%
A_{+}\left(  X_{1}^{2}+X_{1}^{\prime2}\right)  +A_{-}X_{1}X_{1}^{\prime
}\right)  \,,
\end{equation}
where$\,A_{\pm}\equiv\frac{1}{2}\left(  \Omega_{11}\pm\frac{\det
\mathbf{\Omega}}{\Omega_{33}\Omega_{22}-\Omega_{23}^{2}}\right)  $ and $C$ is
a normalization factor. The von Neumann entropy, which measures the
entanglement between the light and atoms, can be calculated as
\cite{PhysRevLett.92.073602}
\begin{equation}
S\equiv-\mathrm{Tr}\left(  \rho\ln\rho\right)  =\frac{\gamma}{e^{\gamma}%
-1}-\ln\left(  1-e^{-\gamma}\right)  \,, \label{ent}%
\end{equation}
where $\gamma\equiv\cosh^{-1}\left(  A_{+}/A_{-}\right)  $. In the limit
$\gamma\ll1$, we have $S\approx1-\ln\gamma$. We calculate $\Delta$ and $S$
numerically and display the results in Fig.~\ref{entropygap}. These two quantities,
unlike the order parameter or $H_{\mathrm{C}}$ which only depends on $x$ and $y$, also depend on
$\lambda
\equiv\omega/\omega_{0}$ like $H_{\mathrm{Q}}$. Therefore the full diagram should be
3-dimensional. In Fig.~\ref{entropygap}, we plot $\Delta$ and $S$ on the $\left(
x,y\right)  $-plane for $\lambda=0.1,\,1,\,10$. Although it is difficult to
distinguish the two phases (normal and superradiant) through $\Delta$ and $S$, the
phase boundary is quite clear in the plots. On the 2nd-order phase transition
boundary, the gap closes and the critical entanglement entropy diverges
logarithmically. By contrast, on the 1st-order phase transition boundary, both
$\Delta$ and $S$ have finite jumps across the phase boundary.

\emph{Critical behavior ---} Let us now turn to the critical behavior of the
tricritical Dicke model. One is often concerned with how the order parameter
behaves near the critical line (i.e., the 2nd-order phase boundary). Consider
a point $(x,y)$ in the superradiance region and close to the critical line, if
we draw a line perpendicular to the critical line through this point and
intercepts the critical line at $(x_{c},y_{c})$, then the order parameter at
$\left(  x,y\right)  $ can be obtained by expanding $f(z)$ in powers of $z$%
\begin{equation}
z^{2}=\frac{2\sqrt{y_{c}^{-2}+4x_{c}^{2}y_{c}^{2}}}{1-\allowbreak5x_{c}%
^{2}\allowbreak\allowbreak}n+o\left(  n\right)  \,,
\end{equation}
where $n$ is the distance between $\left(  x,y\right)  $ and the critical
line. Hence, the critical exponent $\alpha$ defined by $z\propto n^{\alpha}$
is 1/2. However, if the line through $(x,y)$ intercepts the critical line at
the QTP $(x_{tc},y_{tc})$, we have a different scaling:%
\begin{equation}
z^{4}=\frac{5\sqrt{21}}{6}n\,+o\left(  n\right)  , \label{tricritical scaling}%
\end{equation}
which yields an exponent $\alpha=1/4$ for the QTP. In this sense, the
QTP does not belong to the same universality class of the other critical
points in this model, consistent with the general Landau theory of phase
transition.

The critical behavior of the order parameter as described above is determined
by $H_{\mathrm{C}}$. Now let us examine the behavior of the excitation gap $\Delta$
and the entanglement $S$, both of which are governed by $H_{\mathrm{Q}}$. To
this end, we need to find the matrix elements of $\mathbf{\Omega}$. It can be
shown that, on the critical line, $\mathbf{\Omega}$ has eigenvalues $0$,
$\omega_{0}$ and $\sqrt{1+\lambda^{2}}\omega_{0}$. The smallest eigenvalue is
$0$ which indicates that the gap $\Delta$ vanishes, as expected. Furthermore, the entropy $S$
diverges logarithmically according to Eq.~(\ref{ent}). Near the critical line,
to the leading order in $\det\left(  \Omega/\omega_{0}\right)  $, we have
\begin{align}
\Delta/\omega_{0}  &  \sim\left(  1+\lambda^{2}\right)  ^{-1/2}\det\left(
\Omega/\omega_{0}\right)  \,,\label{gap}\\
S  &  \sim1-\frac{1}{2}\ln\left[  \frac{4\left(  \lambda^{2}+1\right)
\det\left(  \Omega/\omega_{0}\right)  }{\lambda^{2}}\right]  \,,
\label{entropy}%
\end{align}
which establishes a universal relation between $S$ and $\Delta$ in the critical
region as
\begin{equation}
S\sim1-\frac{1}{2}\ln\left[  \frac{4\left(  \lambda^{2}+1\right)  ^{3/2}%
\Delta}{\lambda\omega_{0}}\right]  \,. \label{Seps}%
\end{equation}
Equation (\ref{Seps}) represents another key result of this work. Two
important remarks are in order here. First, Eq.~(\ref{Seps}) does not
explicitly contain $z$, which is due to the separation of the classical and
the quantum degrees of freedom aforementioned. The harmonic oscillator modes,
depicted by $H_{\mathrm{Q}}$, are collective modes involving both light and
atoms, emerging above the mean-field ground state of $H_{\mathrm{C}}$ in the
thermodynamic limit, and Eq.~(\ref{Seps}) is solely determined by these modes,
therefore we can call Eq.~(\ref{Seps}) an emergent quantum universality.
Second, Eq.~(\ref{Seps}) is valid near all the critical points despite of the
fact that points around the QTP exhibit different scaling behavior for the
order parameter. It is even valid in the normal phase region below the
critical line where the order parameter vanishes.

Given a point $(x,y)$ sufficiently close to, and a distance $n$ away from, the critical line, the key factor
$\det\left(  \Omega/\omega_{0}\right)  $ in Eq.~(\ref{entropy}) can be
expressed by $n$ as
\begin{equation}
\det\left(  \Omega^{2}/\omega_{0}^{2}\right)  /\lambda^{2}=\beta\sqrt
{y_{c}^{-2}+4x_{c}^{2}y_{c}^{2}}\,n\,+o\left(  n\right)  ,
\end{equation}
where the coefficient $\beta$ takes different values in different critical
regions. If $(x,y)$ is located in the superradiant phase, then $\beta=2$
unless $(x,y)$ approaches the QTP, in which case $\beta=4$. If $(x,y)$ is
located in the normal phase where $z=0$, then $\beta=1$. The scaling exponent
between $\det\left(  \Omega/\omega_{0}\right)  $ and $n,$ is always the same
while the scaling amplitude varies. Consequently, we have $\Delta\propto n^{1/2}$
and the entropy diverges logarithmically in terms of $n$. Another point to
remark is that, as a function of $\lambda$, the critical entanglement entropy
takes the form $S\left(  \lambda\right)  \approx-\frac{1}{2}\ln\left(
\lambda+\lambda^{-1}\right)  +\mathrm{const}$, which indicates that the
entanglement between light and atom is maximized under the resonance condition
$\lambda=1$.

In our model, as in the conventional Dicke model, the strengths of the rotating and the counter-rotating terms are equal. Previous studies have considered a Dicke-type model where these two strengths can have different values and found that there exists a multicritical point in the ground state phase diagram \cite{PhysRevLett.112.173601}. However, in the presence of dissipation, the multicritical point disappears \cite{PhysRevLett.120.183603}. This is related to the disappearance of the superradiance phase in the presence of dissipation when the counter-rotating terms are absent. Due to the presence of the counter-rotating terms, we expect that the QTP in our model should be robust against dissipation. Nevertheless, how the dissipation affect the universal scaling requires further study.

{\em Conclusion ---} In conclusion, we have constructed a generalized Dicke model that supports a
QTP. The phase boundary and the position of the QTP in the parameter space, as
well as the scaling behavior of the order parameter, can be determined from
the mean-field theory and are found analytically. From this, we explicitly
show that the QTP belongs to a different universality class than other points
on the critical line. We further investigated the quantum fluctuations above
the mean-field ground state, and calculated the excitation gap and the
entanglement entropy and their critical behavior near the critical line. We
established a new universal relation between the excitation gap and the
entanglement entropy in the entire critical regime that includes the QTP. The
universality is the result of the separation of the quantum and the classical
degrees of freedom in the thermodynamic limit, being the property of the
emergent collective quantum modes. Our model could be realized using atoms and cavities,
or maybe other platforms, with current technology. Our work opens up new opportunities to investigate quantum tricriticality.

We acknowledge the support from the NSF and the Welch Foundation (Grant No. C-1669).

\end{document}